\documentclass[a4paper, 12pt, oneside]{article}
\usepackage{t1enc}
\usepackage[utf8]{inputenc}
\usepackage[english]{babel}
\usepackage[yyyymmdd]{datetime}
\usepackage[version=4]{mhchem}
\usepackage[top=3cm, bottom=3cm, left=2.5cm, right=2.5cm]{geometry}
\usepackage{amsmath}

\usepackage{float}
\usepackage{multicol}
\usepackage{multirow}
\usepackage{graphicx}
\graphicspath{ {./images/} }
\usepackage{array,multirow,graphicx}
\usepackage{caption}
\usepackage[table]{xcolor}
\usepackage{mhchem}
\usepackage{makecell}
\usepackage{wrapfig}
\usepackage{mathtools}
\usepackage{physics}
\usepackage{changepage}
\usepackage{bm}
\usepackage{subfig}
\usepackage[colorlinks = true,
            linkcolor = black,
            urlcolor  = blue,
            citecolor = black,
            anchorcolor = black]{hyperref}

\title{Zimanyi proceedings}
\author{László Kovács}
\date{March 2023}

\begin{document}
\vspace*{25pt}
\begin{center}
    \LARGE{Charged Kaon Femtoscopy with Lévy Sources in \mbox{$\sqrt{s_{\text{NN}}}=200$ GeV} Au+Au Collisions at PHENIX}
\end{center}

\begin{center}
\vspace{4pt}
\large
    László Kovács\textsuperscript{1} for the PHENIX Collaboration
    
\small
   \textsuperscript{1} Department of Atomic Physics, Faculty of Science, Eötvös Loránd University, Pázmány Péter Sétány 1/A, H-1111 Budapest, Hungary

\end{center}

\begin{center}
\vspace{6pt}
July 18, 2023    
\end{center}

\begin{center}
\vspace{12pt}
\textbf{Abstract}    
\end{center}

\begin{adjustwidth}{30pt}{30pt}
\small \noindent The PHENIX experiment measured  Bose--Einstein quantum--statistical correlations of charged kaons in Au+Au collisions at $\sqrt{s_{\text{NN}}}$ = 200 GeV. The correlation functions are parametrized assuming that the source emitting the particles has a Lévy shape, characterized by the Lévy exponent $\alpha$ and the Lévy scale $R$. By introducing the intercept parameter $\lambda$, we account for the core--halo fraction. The parameters are investigated as a function of transverse mass. The comparison of the parameters measured for kaon--kaon with those measured from pion--pion correlation may clarify the connection of Lévy parameters to physical processes.
\end{adjustwidth}

\newpage
\section{Introduction}
To study the space-time structure of the quark--gluon plasma, the~most commonly used method is femtoscopy. It is a  sub--field of high--energy particles and nuclear physics,  and~it allows us to explore the properties of the matter created in particle collisions on the femtometer scale. Femtoscopy typically investigates correlations of particle pairs. However, it is worth mentioning that prior to the development of femtoscopy, a~similar physical phenomenon was discovered and utilized in the field of radio astronomy. R. Hanbury Brown and R. Q. Twiss measured the angular size of stars by analyzing intensity correlations, which became known as the Hanbury Brown and Twiss effect (HBT) \cite{HanburyBrown:1956bqd}. Roy Glauber's work that laid the foundations of quantum optics~\cite{PhysRevLett.10.84,RevModPhys.78.1267,GLAUBER20063} greatly increased our understanding of this effect. G. Goldhaber and his collaborators observed intensity correlations among same-charged pions while searching for $\rho$ mesons in high-energy collisions. These correlations were explained by G. Goldhaber, S. Goldhaber, W-Y. Lee, and~A. Pais (GGLP), based on the Bose--Einstein symmetrization of the wave-function of identical pion pairs~\cite{PhysRev.120.300}. This is the reason why these correlations are often called Bose--Einstein correlations. Due to the relationship between the two--particle Bose--Einstein correlation function and the phase-space density of the particle-emitting source, by~measuring the correlation function we can obtain information about the source~function. 

Let us note that while there are conceptual similarities between the HBT effect in radio astronomy and the correlations studied in femtoscopy, there are also fundamental differences. In~femtoscopy, we can extract information about the space-time structure of the particle source, often represented by femtoscopic radii. On~the other hand, the~HBT effect in radio astronomy provides insights into the spectral-angular structure of the source radiation. 

The fundamentals of modern correlation femtoscopy were established by Kopylov and Podgoretsky~\cite{Kopylov:1974th,Podgoretsky:1989th}. They successfully overcame the drawback of the GGLP technique by utilizing the momentum differences of the particle pairs instead of the opening~angles.

Based on the central limit theorem, it is a good approach to assume a Gaussian shape for the phase-space density of the particle--emitting source. However, we can go further and take a more general approach. Anomalous diffusion indicates the appearance of Lévy--stable distributions for the source~\cite{Csorgo:2003uv,PhysRevLett.82.3563}. In~Ref.~\cite{PhysRevC.97.064911}, it was found that Lévy--stable source distributions in $\sqrt{s_{\text{NN}}}=200$ GeV Au+Au collisions give a high--quality, statistically acceptable description of the measured correlation functions in the case of pions. In~the present paper, we will investigate kaon--kaon correlation functions assuming a Lévy shaped source. By~comparing the obtained results with the pion data, more insights can be gained regarding the Lévy~parameters. 

\begin{figure}[H]
\centering
\includegraphics[width=0.5\textwidth]{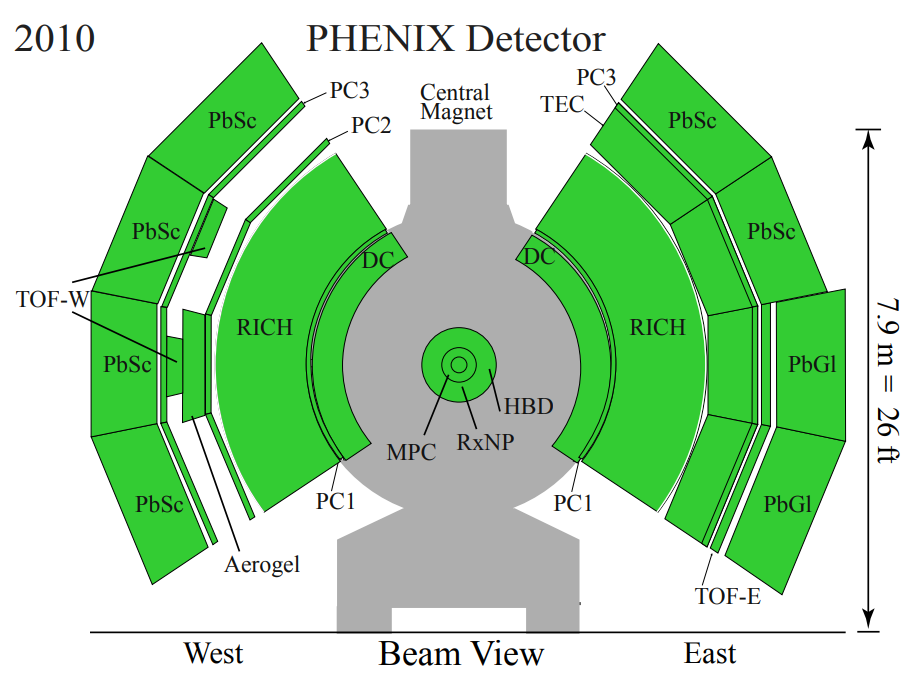}
\caption{View of the PHENIX central arm spectrometer detector setup in the 2010 data-taking~period.}
\label{fig:det}  
\end{figure}

The dataset used in this analysis is Au+Au collisions at $\sqrt{s_{\text{NN}}}=200$ GeV recorded by the PHENIX (\textbf{P}ioneering \textbf{H}igh \textbf{E}nergy \textbf{N}uclear \textbf{I}nteraction e\textbf{X}periment) detector. It is one of the four experiments that have taken data at the relativistic heavy ion collider (RHIC) in Brookhaven National Laboratory. Its primary mission was to search for a new state of matter called the quark--gluon plasma and~to study various different particle types produced in heavy ion collisions, such as photons, electrons, muons, and~charged hadrons. A~beam view layout of the PHENIX detector can be seen in Figure~\ref{fig:det}. The~detectors can be divided into four main~subgroups:
\begin{enumerate}
    \item Global detectors characterize the nature of heavy ion collision events, i.e.,~zero degree calorimeters (ZDC) and beam-beam counters (BBC);
    \item Mid-rapidity detectors form the ``central arm spectrometer'', which consists of three sets of pad chambers (PC),~drift chambers (DC), electromagnetic calorimeters (EmCal), and~time--of--flight detectors (ToF), are used for energy, momentum, and~mass measurements;
    \item Two muon spectrometers at forward rapidity;
    \item A triggering and computing system to select and archive events of potential physics interest.
\end{enumerate}

\section{Femtoscopy and Lévy~Sources}
As we mentioned in the previous section, there is a connection between the Bose--Einstein correlation function and the phase--space density of the particle-emitting source. Let us discuss this relationship in more detail. The~one-- and two--particle momentum distributions can be expressed as~\cite{Yano:1978gk}
\begin{equation}
    N_1(p) = \int d^4 r S(r,p) |\psi_{p}(r)|^2,
    \label{eq:mom_dists1}
\end{equation}
\begin{equation}
    N_2(p_1,p_2) = \int d^4 r_1 d^4 r_2 S(r_1, p_1) S(r_2, p_2)|\psi_{p_1,p_2}^{(2)}(r_1,r_2)|^2,
    \label{eq:mom_dists2}
\end{equation}
where $S(r,p)$ is the source function, which describes the probability density of particle creation at space-time point $r$ with four-momentum $p$; $\psi_{p}(r)$ denotes the single-particle wave function; and~$\psi_{p_1,p_2}^{(2)}$ is the two--particle wave function, which must be symmetric in the spatial variables $r_1$ and $r_2$ for bosons. Bose--Einstein correlations arise from this symmetrization effect. Using Equations \eqref{eq:mom_dists1} and \eqref{eq:mom_dists2}, we can express the two--particle correlation function as~\cite{Csorgo:1999sj,Wiedemann:1999qn}
\begin{align}
C_2(p_1,p_2) = \frac{N_2(p_1,p_2)}{N_1(p_1)N_1(p_2)}.
\end{align}

Let us introduce the average momentum $K = 0.5(p_1+p_2)$ and relative momentum $q = p_1-p_2$ as new variables. If~$p_1 \approx p_2 \approx K$ and the final state interactions are neglected, the~two--particle correlation function can be written as
\begin{equation}
C_2^{(0)}(q,K) \approx 1 + \frac{|\tilde{S}(q,K)|^2}{|\tilde{S}(0,K)|^2},
\label{eq:C2C}
\end{equation}
where the superscript (0) denotes the neglection of final state interactions and $\tilde{S}(q,K)$ is the Fourier transform of the source with
\begin{equation}
\tilde{S}(q,K) = \int S(x,K) e^{iqx} d^4x.
\end{equation}

The significance of Equation \eqref{eq:C2C} lies in the fact that by measuring the Bose--Einstein correlation function, we can obtain information about the spatial shape of the source function.

The correlation function depends on the four-momentum difference and the average four-momentum. Since the Lorentz product of  $q$ and $K$ is zero in the case of identical particles, the~correlation function depends only on the spatial \textbf{q} instead of the four dimensional $q$ vector:
\begin{align}
q K = q_0 K_0 - \textbf{qK} = 0 \hspace{0.3cm} \Rightarrow \hspace{0.3cm} q_0 = \frac{\textbf{qK}}{K_0}.
\end{align}

In Ref.~\cite{Podgoretsky:1989th}, Kopylov and Podgoretsky showcased the three-dimensional character of the momentum correlation effect due to the particles detected~on-mass-shell.

For our correlation function measurements, we use the co-moving system (LCMS) frame, where it was found in earlier measurements~\cite{PhysRevLett.93.152302} that the correlation function is nearly spherically symmetric. Due to the relatively low number of produced kaons and this symmetrical characteristic, we have chosen to perform a one-dimensional (1D) analysis instead of a three-dimensional (3D) one. Based on Ref.~\cite{PhysRevC.97.064911}, we used $Q = |q_\textmd{LCMS}|$ as the 1D variable of the correlation~function.

We assumed that the source emitting the particles has a Lévy shape. The~symmetric Lévy--stable distribution is defined as
\begin{align}
\mathcal{L}(\textbf{r};R,\alpha) = \frac{1}{(2\pi)^3} \int \text{d}^3 \bm{\varphi} \:  e^{i\bm{\varphi}\textbf{r}} e^{-\frac{1}{2}|\bm{\varphi}R|^\alpha},
\end{align}
where $R$ is the Lévy scale parameter, $\alpha$ is the Lévy exponent, and~$\bm{\varphi}$ is a three-dimensional integration variable. The~$\alpha$ parameter describes the shape of the distribution, in~the case of a Gaussian distribution $\alpha$ = 2, while for a Cauchy distribution, the~value of $\alpha$ is~1. 

In case of such Lévy--stable source functions, the~raw (i.e., final-state
interaction neglected) correlation function in the observable $Q$-range will be~\cite{Csorgo:2003uv}
\begin{align}
C_2^{(0)}(Q;\lambda,R,\alpha)=1+\lambda e^{-|QR|^\alpha},
\label{eq:levy_corr_func}
\end{align}
where the intercept parameter $\lambda$ is introduced as the extrapolated $C_2^{(0)}(Q=0)$ value. 

According to the core–halo model~\cite{Lednicky:1979ig,Csorgo:1994in}, the~source can be divided into two parts:  the core, which contains the promptly produced particles, and~ the halo, which is composed of the products of resonance decays. The~ratio of these two parts can be characterized by the correlation strength parameter:
\begin{align}
\lambda = \left( \frac{N_\textmd{core}}{N_\textmd{core}+N_\textmd{halo}} \right)^2,
\end{align}
where $N_\textmd{core}$ refers to the number of particles produced in the core, while $N_\textmd{halo}$ denotes the number of particles produced in the halo. Considering that the particles from the halo contribute to the correlation function as an unresolvably narrow peak around $Q$ = 0, we indeed see that this $\lambda$ value will be the extrapolated $C_2^{(0)}(Q=0)$ value.

For charged particles, the~most significant final state interaction is the Coulomb interaction. To~take care of this effect, we used the Sinyukov–Bowler method~\cite{Sinyukov:1998fc,1991PhLB27069B}. Taking an additional possible linear background shape into account, our final assumption for the functional form of the correlation function is
\begin{align}
C_2(Q;\lambda,R,\alpha) = \left[ 1 - \lambda + K(q_\textmd{inv};\alpha,R) \cdot \lambda  \cdot \left( 1 + e^{-| Q R |^{\alpha}} \right) \right] \cdot N \cdot (1+\epsilon Q),
\label{eq:Bowler_Sinyukov}
\end{align}
where $K$ is the Coulomb correction, $N$ is the normalization parameter, and~$\epsilon$ represents a small background long-range correlation effect. Let us note that the Coulomb correction is a function of $q_\textmd{inv}$, which is the Lorentz invariant four-momentum difference\footnote{This variable can be expressed in the PCMS system, which is the pair rest frame: $q_\textmd{inv} = |\textbf{q}_\textmd{PCMS}|$.}, while the correlation function has a different variable, denoted by $Q$. We calculated the Coulomb correction $K$ with the variable $Q$ and analyzed the error coming from this approximation, which was handled as a source of systematic uncertainty the same way as in Ref.~\cite{PhysRevC.97.064911}. The~correction is quite small compared to other sources, so it does not mean a large additional term to the systematics, and~this way we have more comparable results to the~pions.

\section{Motivation}
In Ref.~\cite{PhysRevC.97.064911}, the~significance of the appearance of the Lévy distribution in the case of pion–pion correlations was investigated. In~order to dive into the exploration of the L\'evy-shape, we aimed to analyze kaon~correlations.

The Lévy source parameters for kaon–kaon two particle correlations have never been measured in PHENIX~before.

Anomalous diffusion could be a reason for a Lévy distribution~\cite{Csanad:2007fr}. In~such a scenario, the~Lévy index for the different particles is different, i.e.,~$\alpha_\textmd{L\'evy}^\pi  \neq \alpha_\textmd{L\'evy}^{\textmd{K}}$. The~smaller the cross section is, the~longer the mean free path, thus the longer the power-law like ``tail'' of the source distribution. Kaons have a smaller cross section than pions, so we would expect that $\alpha_\textmd{L\'evy}^\pi  > \alpha_\textmd{L\'evy}^{\textmd{K}}$. If~this explanation fails to match the reality, other alternatives should be~investigated.

The Lévy width (or, equivalently, scale parameter) $R$ has an unclear interpretation. It exhibits similar behavior as the Gaussian source radii, but~its precise relation to the geometrical source size is not clear. By~measuring this parameter for kaons, we can get closer to clarifying its precise physical~interpretation.

\section{Measurement~Details}
In this measurement, we analyzed Au+Au collisions at $\sqrt{s_\textmd{NN}}=$ 200 GeV. The~dataset consists of about 7.3 billion (minimum bias triggered) events. As~the number of produced kaons is relatively low, all minimum bias events (of any centrality) were taken together. We cut out those events whose distance from the nominal collision point was greater than \mbox{30 cm} along the beam~axis.

We also need to take into account the detector inefficiencies and the particularities of the track reconstruction algorithm, which sometimes splits one track into two. On~the other hand, when two different tracks are too close to each other, it is possible that they will be detected as a single track. To~remove these possible effects, we applied cuts in the $\Delta \phi$ and $\Delta z$ variables, where $\Delta \phi$ and $\Delta z$ stand for the azimuthal angle and longitudinal position difference of track pairs, respectively (as measured in the drift chamber, the~main tracking detector).

\begin{figure}[H]
\centering
\captionsetup[subfigure]{justification=centering}
  \subfloat[]{\includegraphics[width=0.49\textwidth]{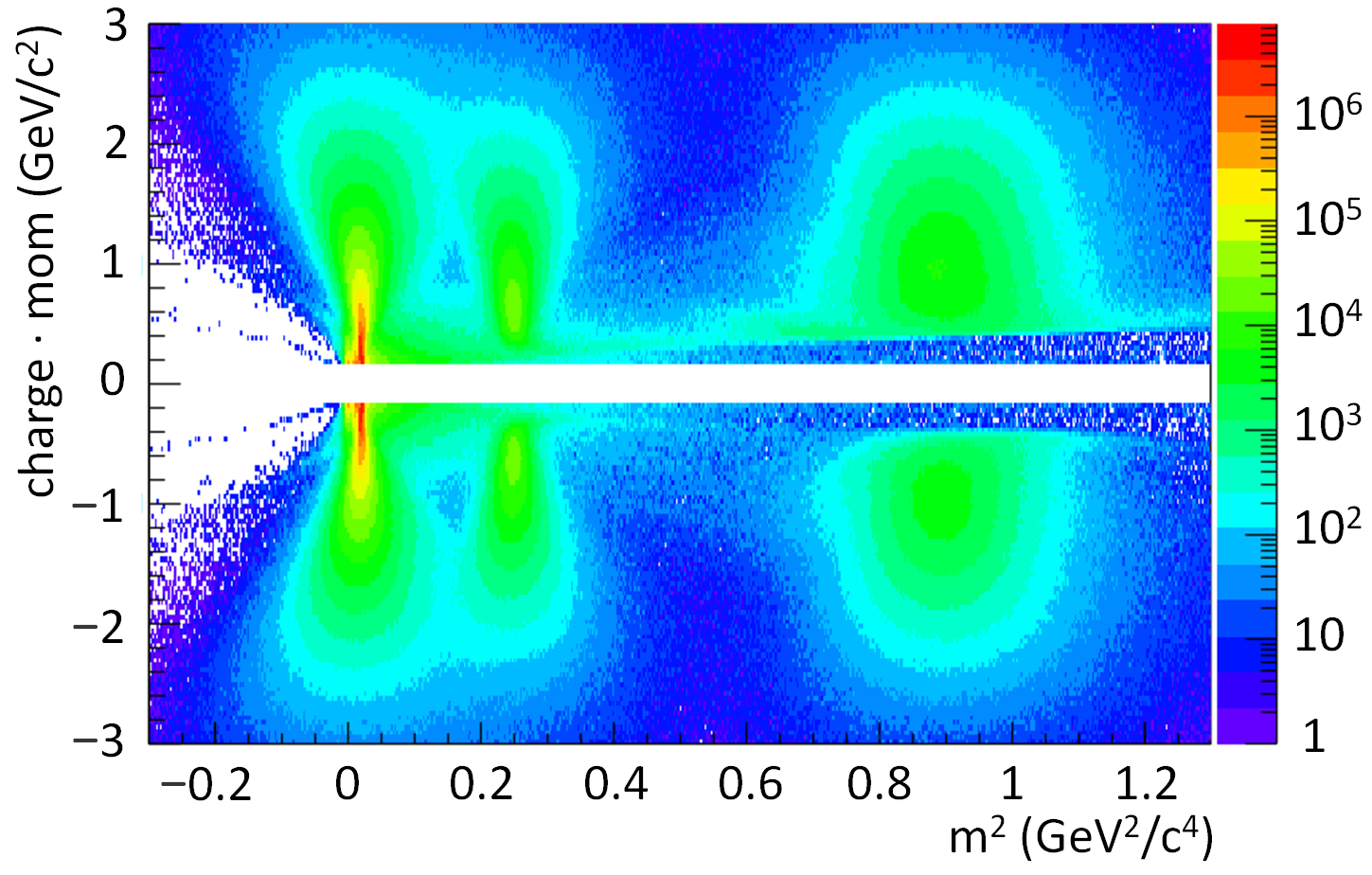}\label{fig:PID_1}}
  \hfill
  \subfloat[]{\includegraphics[width=0.49\textwidth]{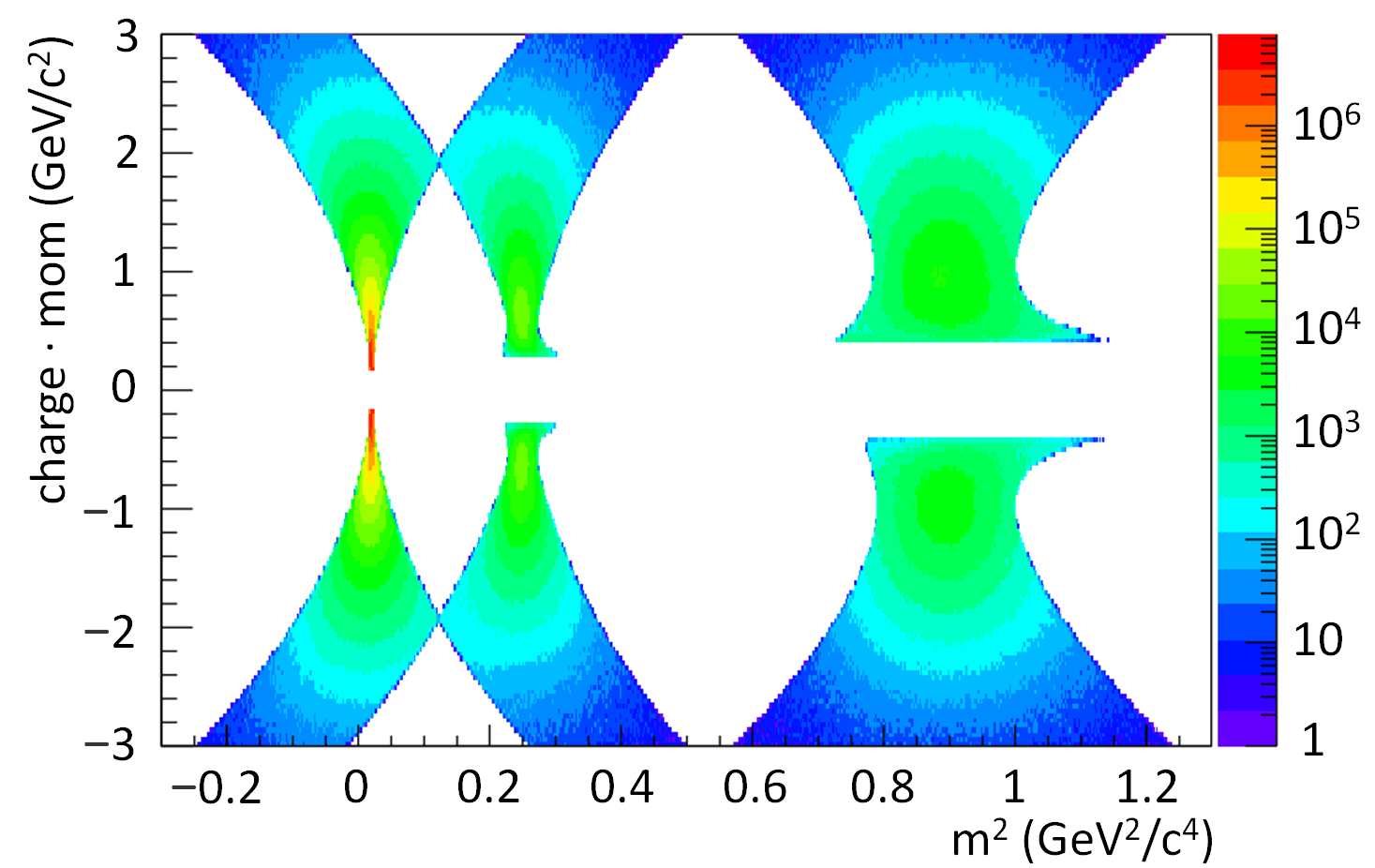}\label{fig:PID_2}}
  \caption{Example plots for~PID. (\textbf{a}) Scatter plot of the charge times momentum vs. $m^2$ in TOF West with no cut. (\textbf{b}) Scatter plot of the charge times momentum vs. $m^2$ in TOF West after the applied cuts.}
  \label{fig2}
\end{figure}

For particle identification (PID), we calculated the square of the particle mass:
\begin{equation}
    m^2 = \frac{\textbf{p}^2}{c^2} \left[ \left(\frac{ct}{L}\right)^2-1 \right],
\end{equation}
where $t$ is the time of flight (measured either in the PbSc or the ToF East/West detectors), $L$ is the path length, and \textbf{p} is the momentum. The~distributions of $m^2$ were fitted mostly using single Gaussians; in cases of merging peaks, a double Gaussian was applied. To~identify kaons, we applied a 2.5 standard deviation ($\sigma_\text{p}$) cut around the nominal kaon $m^2$ peak position and~a 2.5$\sigma_\text{p}$ veto cut around the pion and proton $m^2$ peaks. An~example scatter plot of the charge times momentum vs. $m^2$ before and after the cuts can be seen in Figure~\ref{fig2}a,b, respectively. 
\vspace{-12pt}

Since the correlation function's dependence on $K$ is smoother than its dependence on $Q$, it is reasonable to create several $K$  bins, and~in each bin, the~$Q$ dependence can be investigated. At~midrapidity, the~transverse mass $m_\textmd{T}$ can be used instead of $K$:
\begin{equation}
    m_\textmd{T} = \sqrt{m^2 + (K_\textmd{T}/c)^2},
\end{equation}
where $m$ is the mass of the particle and~\begin{equation}
    K_\textmd{T} = \sqrt{K_\textmd{x}^2 + K_\textmd{y}^2}
\end{equation}
is the average transverse momentum. In~this analysis, 7 $m_\textmd{T}$ bins were created. In~each bin, we analyzed the dependence of the correlation function on $Q$. 

To measure the correlation functions, ``actual'' (foreground) and ``background'' distributions of the kaon pairs were created. To~construct the actual pair distribution, the~momentum differences were calculated of the same-charged particles from the same event and~filled into a histogram. Since there are other effects (stemming from acceptance, single-particle distributions, efficiency, etc.) in the actual pair distribution that are not related to the HBT effect, we have to cancel them out with a properly constructed background distribution that contains pairs from different events, where there can be no HBT effect. To~create the background pair distribution, we employed the same mixed event method as described in Ref.~\cite{PhysRevC.97.064911}. The~first step is to construct a pool that contains several events. This pool needs to be at least as large as the number of produced kaons in the event with the highest multiplicity. In~order to ensure that we meet this condition, a~pool with \mbox{50 events} was used. Every time we process an event to construct the actual pair distribution, we construct a mixed event for the background distribution as well. Since we do not want to introduce any correlations and we would like to avoid the presence of the quantum-statistical correlation between the particles in the background, we have to select the particles as follows. First of all, to~ensure that the background event exhibits the same kinematics and acceptance effects, we have to construct the background event from events of similar centrality and with a similar $z$ coordinate of the collision vertex. To~accomplish this, we used 5\% wide centrality and 2 cm wide $z$-vertex bins. Secondly, it is essential that the selected particles for the background pair distribution originate from different events. After~the particle selection from mixed eventsm we calculate the momentum differences of these~particles.

The two--particle correlation function can be calculated from the ratio of the normalized actual and background pair distributions:
\begin{align}
C_2(m_\textmd{T},Q) = \frac{A(m_\textmd{T},Q)}{B(m_\textmd{T},Q)} \cdot \frac{\int_{Q_\textmd{min}}^{Q_\textmd{max}} B(m_\textmd{T},Q)}{\int_{Q_\textmd{min}}^{Q_\textmd{max}} A(m_\textmd{T},Q)},
\end{align}
where $A$ is the actual, $B$ is the background pair distribution, and~the integral is performed over a range ($Q_\textmd{min}-Q_\textmd{max}$), where the correlation function does not exhibit quantum statistical~features.

\begin{figure}[H]
\centering
\includegraphics[width=0.8\textwidth]{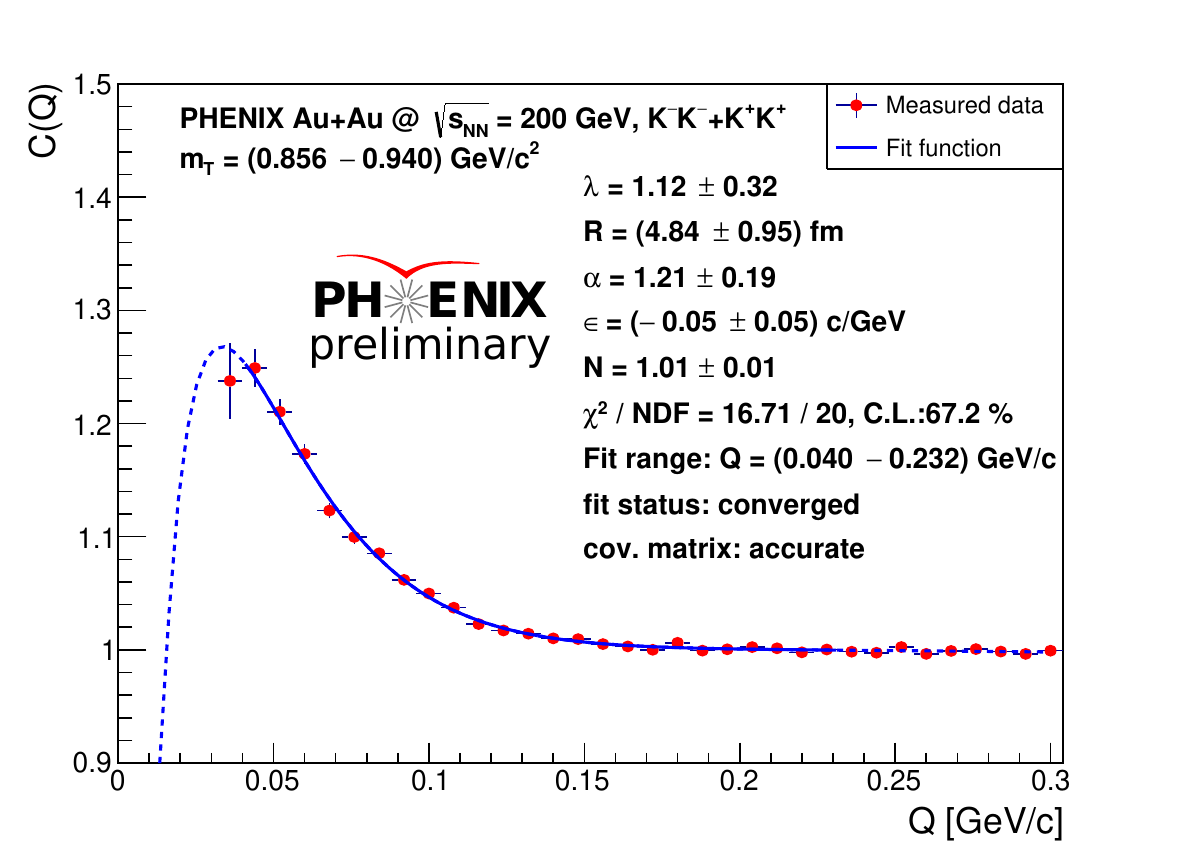}
\caption{Anexample fit with the Coulomb-corrected correlation function based on a Lévy source for kaon pairs with transverse mass ranging from 0.856 GeV/$c^2$ to 0.940 GeV/$c^2$.}
\label{fig:fit} 
\end{figure}

We fitted the measured correlation functions with the Coulomb-corrected L\'evy-type correlation function and the linear background. Of~all the final state interactions, the~Coulomb effect has the greatest impact as~it causes same-charged pairs to repel each other. As~shown in Figure~\ref{fig:fit}, the~function drops off sharply at small $Q$ due to the Coulomb effect.

To deal with the Coulomb correction, we applied the same method that was used in Ref.~\cite{PhysRevC.97.064911} for pions; we did this to obtain comparable results. First, we needed to numerically solve the Coulomb correction; then, the results were loaded into a binary look up table as described in Refs.~\cite{Csanad:2019lkp,Csanad:2019cns}. This numerical table contains the values discretely, so interpolation was needed, which can cause numerical fluctuations. These fluctuations can be handled with a proper iterative fitting procedure. The~first round fit was performed with a functional form of the correlation function incorporating the Coulomb correction, which yields a set of parameters  $\lambda_0$, $R_0$, and $\alpha_0$. Based on Equation \eqref{eq:Bowler_Sinyukov}, the~second round fit was with
\begin{equation}
C_2^{(0)}(Q;\lambda,R,\alpha) \frac{C_2(Q;\lambda_\textmd{n},R_\textmd{n},\alpha_\textmd{n})}{C_2^{(0)}(Q;\lambda_\textmd{n},R_\textmd{n},\alpha_\textmd{n})} \cdot N \cdot (1+\epsilon Q),
\end{equation}
where the fitted parameters are denoted as $\lambda$, $R$, $\alpha$, $N$, and~$\epsilon$. In~the second round, the~values of $\lambda_\text{n}$, $R_\text{n}$, and~$\alpha_\text{n}$ were equal to the corresponding values from the first fit. The~correlation function without the Coulomb correction is denoted by $C_2^{(0)}(Q;\lambda,R,\alpha)$, while $C_2(Q;\lambda,R,\alpha)$ refers to the Coulomb-corrected one. We continued this iterative procedure until the parameters of the previous fit ($\lambda_\textmd{n},R_\textmd{n},\alpha_\textmd{n}$) and the new ones from the latest fit ($\lambda_\textmd{n+1},R_\textmd{n+1},\alpha_\textmd{n+1}$) differed less then 2\%. Let us note that usually $N\approx1$ and $\epsilon\approx0$, and~these parameters converge faster than $\lambda$, $R$ and $\alpha$, so only the latter parameters are used in the test of the convergence~criteria.

To determine the systematic uncertainties of the parameters, alternative measurement settings were applied. These considered settings can be seen in Table~\ref{tab:syst}. In~case of the PID cut, the~default setting was 2.5$\sigma_\text{p}$, while the lower one was 2.0$\sigma_\text{p}$ and~the upper one was 3.0$\sigma_\text{p}$. As~for the PC3 matching cut, there was no cut in the default setting, but~an alternative cut of 2.0$\sigma_\text{m}$ was applied, where $\sigma_\text{m}$ is the standard deviation of the differenece of the projected track position and the closest hit position in the detector, in~both the $\phi$ and $z$ directions.  Regarding the EMCal/ToF track matching cut, the~default setting was 1.5$\sigma_\text{m}$, while the lower one was 1.0$\sigma_\text{m}$ and~the upper one was 3.5$\sigma_\text{m}$. Pair cuts were applied in the $\Delta \phi-\Delta z$ plane by cutting off a two-dimensional region, as~described in Ref.~\cite{PhysRevC.97.064911}. The~fit range ($Q_\text{min}-Q_\text{max}$) was also varied. We modified the default setting with $\pm$8 MeV/$c$. As~we mentioned earlier in this paper, the~Coulomb correction is a function of $q_\text{inv}$, while the correlation function has a different variable, denoted by $Q$. We calculated the Coulomb correction with the variable $Q$ and treated this approximation as a systematic uncertainty. The~parameters were recalculated by individually changing each of the measurement settings; then, we calculated the relative difference of the values of the parameters obtained from the alternative settings and from the default settings. After~calculating the relative differences for all the settings, we obtained the final systematics by taking into account the statistical uncertainties of the data points. A~similar argument can be found in Ref.~\cite{Barlow:2002yb}. 
\begin{table}[H]
\centering
\begin{tabular}{cc}
\hline
Setting name               & Settings       \\ \hline
PID cut                    & 3 cut settings \\
PC3 matching cut           & 1 cut setting  \\
EMCal/ToF matching cut     & 3 cut settings \\
DC pair cut                & 3 cut settings \\
ToF East pair cut          & 3 cut settings \\
ToF Wast pair cut          & 3 cut settings \\
EMCal pair cut             & 3 cut settings \\
Fit range ($Q_\text{max}$) & 3 ranges       \\
Fit range ($Q_\text{min}$) & 3 ranges       \\
Coulomb correction variable             & 2 versions     \\ \hline
\end{tabular}
\caption{\label{tab:syst}The varied settings in order to determine the systematic uncertainties of the results.}
\end{table}

The method we used is described below. The~variance of the difference of two \mbox{variables is}
\begin{align}
\sigma^2(A_\textmd{def} - A_\textmd{alt}) = \sigma^2(A_\textmd{def}) + \sigma^2(A_\textmd{alt}) - 2\textmd{cov}(A_\textmd{def},A_\textmd{alt}),
\end{align}
where $A_\textmd{def}$ represents the parameter value obtained from the default cut, while $A_\text{alt}$, the parameter value, is obtained from the alternative cut; $\textmd{cov}(A_\textmd{def},A_\textmd{alt})=\rho\sigma(A_\textmd{def})\sigma(A_\textmd{alt})$ is the covariance matrix; and $\rho$ is the correlation coefficient. The~total uncertainties are composed of the systematic and the statistical uncertainties:
\begin{align}
\sigma_\textmd{tot}^2 = \sigma_\textmd{stat}^2 + \sigma_\textmd{syst}^2 \hspace*{2cm} \textmd{so} \hspace*{2cm} \sigma_\textmd{syst}^2 = \sigma_\textmd{tot}^2 - \sigma_\textmd{stat}^2. 
\end{align}

We require that the total uncertainty cover 1 standard deviation (1$\sigma$), i.e.,~$\sigma_\textmd{tot}=|A_\textmd{def}-A_\textmd{alt}|$. Thus,
\begin{align}
\sigma_\textmd{syst}^2 = (A_\textmd{def}-A_\textmd{alt})^2 -\sigma_\textmd{stat}^2(A_\textmd{def}) - \sigma_\textmd{stat}^2(A_\textmd{alt}) + 2\rho\sigma_\textmd{stat}(A_\textmd{def})\sigma_\textmd{stat}(A_\textmd{alt})
\label{eq:new_syst_estim}.
\end{align}

The advantage of using Equation \eqref{eq:new_syst_estim} is that it allows us to consider the impact of the statistical uncertainties. In~this analysis, we assumed that $A_\textmd{def}$ is completely correlated with $A_\textmd{alt}$, thus $\rho=1$. The~final systematic uncertainties were obtained by taking the squared sum of the calculated $\sigma_{\textmd{syst}}$ values for each alternative~setting.

\section{Results}
In this section, we present our main results: a comparison of the transverse mass dependence of the Lévy parameters in the case of kaon–kaon and pion–pion~correlations.

\begin{figure}[H]
\centering
\includegraphics[width=0.8\textwidth]{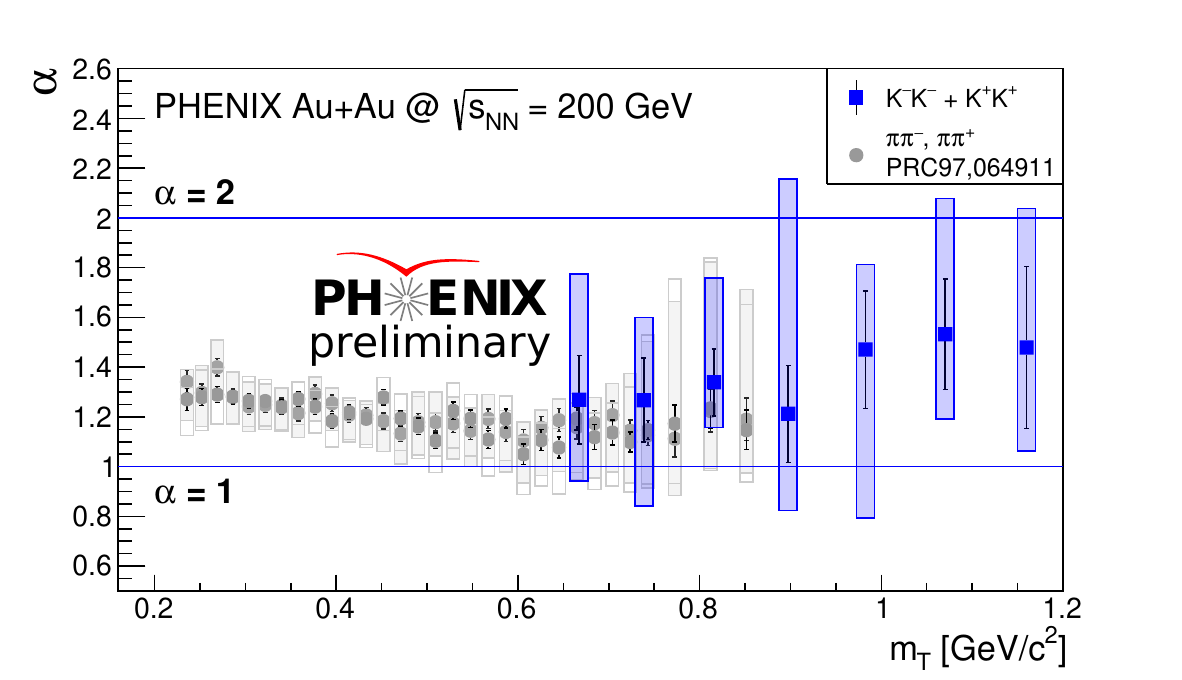}
\caption{Values of the $\alpha$ parameter in the case of pions and kaons. Boxes indicate the systematic uncertainties, while error bars are used to represent the statistical~ones.}
\label{fig:alpha} 
\end{figure}

One of the main reasons why analyzing the kaon–kaon Lévy distribution is interesting is because it could shed light on the physical interpretation of the Lévy exponent. The~Lévy exponent $\alpha$ is shown in Figure~\ref{fig:alpha}. Within~statistical uncertainties, we can draw the conclusion that the value of the parameter is between 1 and 2; however, the~systematic uncertainties are quite large. As~it is described in Ref.~\cite{Csanad:2007fr}, a~higher $\alpha$ value is expected for pions than for kaons based on the anomalous diffusion; however, we cannot observe this trend here, which indicates that beside anomalous diffusion of hadrons, there may be other physical processes causing the appearance of Lévy distributions, such as the resonances, as was concluded in Ref.~\cite{lednicky1992influence}. The~violation of the $m_\text{T}$-scaling of the two-pion and two-kaon correlations suggested by hydrodynamic models was explained by a rescattering phase in Ref.~\cite{ALICE:2017iga}, which was not taken into account in the pure hydrodynamic models. A~slight increase in the $\alpha$ values of the kaon measurements can be observed, although~the large uncertainties do not allow us to draw any strong~conclusions.

\begin{figure}[H]
\centering
\includegraphics[width=0.8\textwidth]{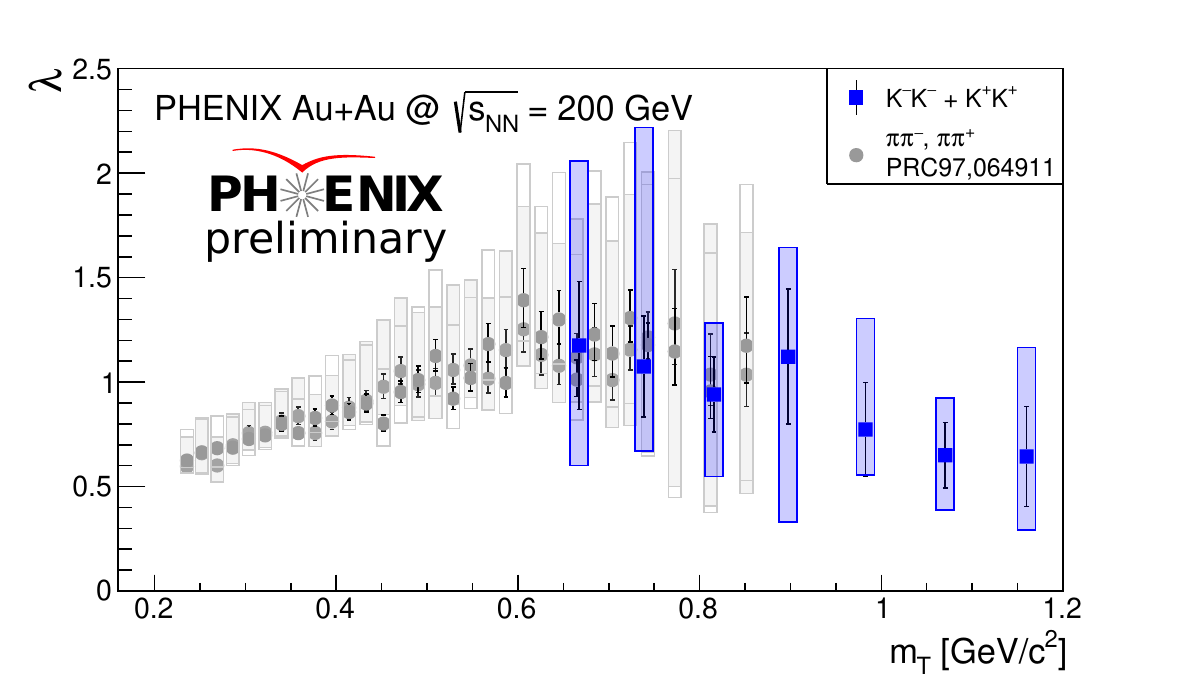}
\caption{Values of the $\lambda$ parameter in the case of pions and kaons. Boxes indicate the systematic uncertainties, while error bars are used to represent the statistical~ones.}
\label{fig:lambda} 
\end{figure}

The transverse mass dependence of the intercept parameter $\lambda$ is shown in Figure~\ref{fig:lambda}. Compared to the pion data, it is not inconsistent with the given large uncertainties, having approximately matching values at around $m_\textmd{T}$ = 0.7 GeV/$c^2$, although~their trends appear to be different. No significant $m_\textmd{T}$ dependence was observed for this parameter, and it is fairly constant; however, we have to note that a slight decreasing trend is visible. This parameter characterizes the strength of the correlation as it was introduced as the extrapolated $C_2^{(0)}(Q=0)$ value. Since there is no correlation between particles of different species, a~possible worsening of PID efficiency may cause a decrease in the value of this parameter as~our dataset may contain particles other than~kaons.

\begin{figure}[H]
\centering
\includegraphics[width=0.8\textwidth]{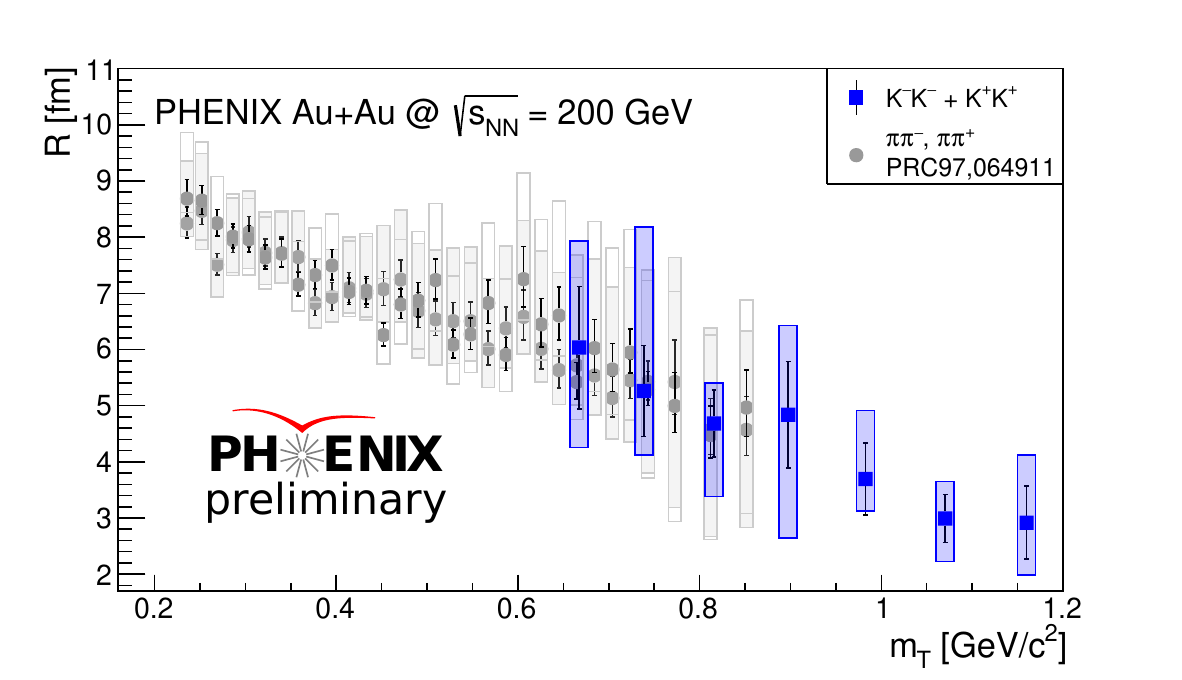}
\caption{Values of the $R$ parameter in the case of pions and kaons. Boxes indicate the systematic uncertainties, while error bars are used to represent the statistical~ones.}
\label{fig:R} 
\end{figure}

The transverse mass dependence of the Lévy scale parameter $R$ is shown in Figure~\ref{fig:R}. The~$m_\text{T}$ scaling of HBT radii across particle species has been predicted in Ref.~\cite{Csanad:2008gt}. From~theoretical works, e.g.,~Refs.~\cite{Csorgo:1999wx, Csorgo:2003uv}, we know that $R$ is not an RMS so it cannot be related to the source size directly. However, it was clear from previously published analyses (see Refs.~\cite{PhysRevC.97.064911, Lokos:2018qdl, Lokos:2018dqq, Kurgyis:2020vbz, Kincses:2019rug, Kincses:2017zlb, Kurgyis:2018zck}) that the Lévy-scale $R$ exhibits a similar trend as its Gaussian counterpart, namely, it decreases with $m_\textmd{T}$. In~the case of a Gaussian source, hydrodynamic models predict a linear scaling for its inverse square~\cite{PhysRevLett.74.4400,Makhlin:1987gm,PhysRevC.54.1390}:
\begin{equation}
    \frac{1}{R^2} = A \cdot m_\textmd{T} + B.
\end{equation}

\begin{figure}[H]
\centering
\includegraphics[width=0.8\textwidth]{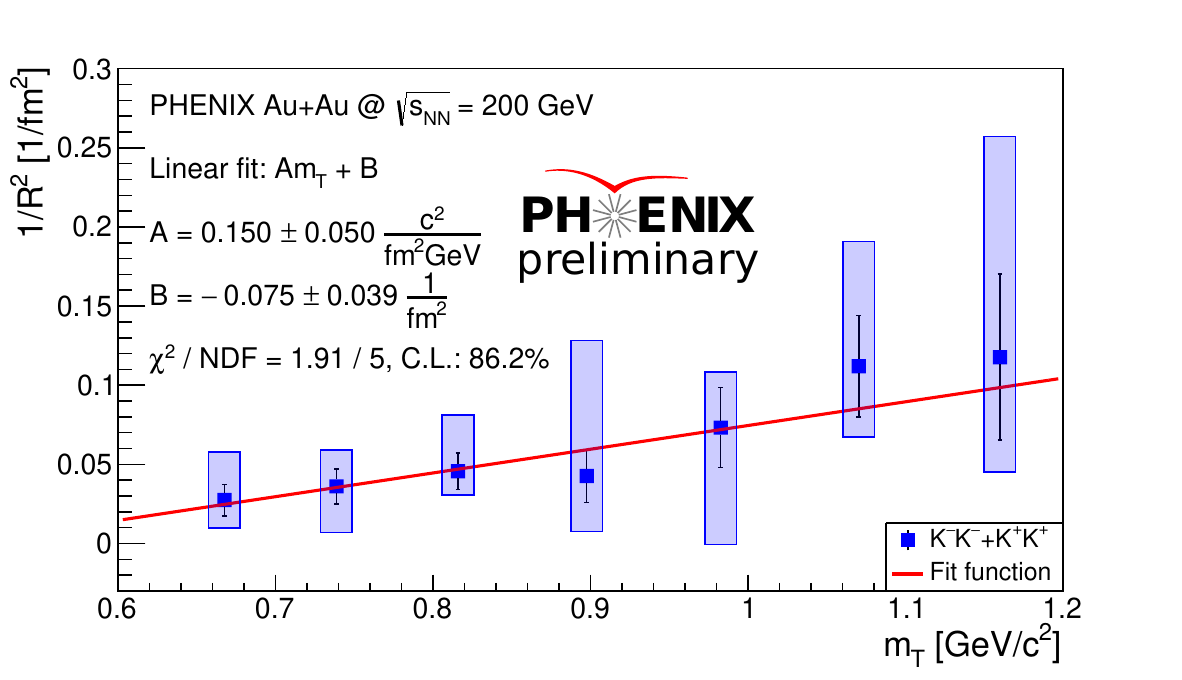}
\caption{The transverse mass dependence of the 1/$R^2$ points. It is worthwhile to note that due to the large uncertainties, one could fit these data points with different powers of $m_\text{T}$ as well. A~line is fitted to the data points, and~the fitted parameters are shown in the legend. Boxes indicate the systematic uncertainties, while error bars are used to represent the statistical~ones.}
\label{fig:Rsq} 
\end{figure}

As we see it on Figure~\ref{fig:Rsq}, the~linear scaling holds for the Lévy source as well, requiring the need for further theoretical~investigations.

It is worthwhile to note that there is a significant amount of point-by-point fluctuation in the systematic uncertainties. Furthermore, the~non-fluctuating part of the systematic uncertainty of the pion and kaon data points is also partly~correlated.

In Ref.~\cite{PhysRevC.97.064911}, a~new empirical scaling variable was found:
\begin{align}
    \widehat{R} = \frac{R}{\lambda (1 + \alpha)}.
\end{align}

\begin{figure}[H]
\centering
\includegraphics[width=0.8\textwidth]{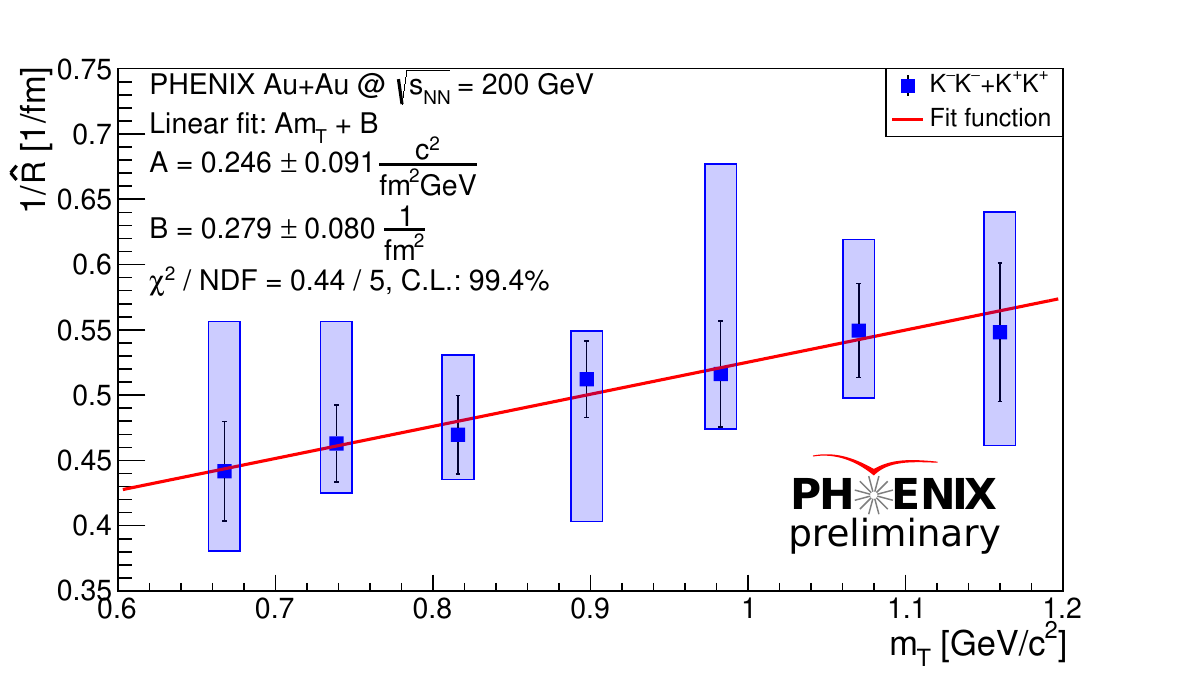}
\caption{The transverse mass dependence of the 1/$\hat{R}$ points. It is worthwhile to note that due to the large uncertainties, one could fit these data points with different powers of $m_\text{T}$ as well. A~line is fitted to the data points, and~the fitted parameters are shown in the legend. Boxes indicate the systematic uncertainties, while error bars are used to represent the statistical~ones.}
\label{fig:Rhat} 
\end{figure}

The motivation behind this parameter was the fact that the $\alpha$, $R$, and~$\lambda$ parameters are strongly correlated, and~it is possible to obtain good fits with multiple sets of co-varied parameters. The~discovery of $\widehat{R}$ was made without any theoretical motivation, and~in Ref.~\cite{PhysRevC.97.064911} it was observed that $\frac{1}{\widehat{R}}$ scales linearly with $m_\textmd{T}$. In~Figure~\ref{fig:Rhat}, we can see the same linear behavior for kaons as~well.

\section{Conclusions}
In this paper, we discussed two-kaon Bose--Einstein correlation functions in Au+Au collisions at $\sqrt{s_\textmd{NN}}=$ 200 GeV, from~the PHENIX experiment. We assumed that the source has a Lévy shape. The~Lévy parameters were investigated as functions of $m_\textmd{T}$ and~compared to the pion results. In~the case of the Lévy stability index $\alpha$, the~large uncertainties prevent us from drawing any strong conclusions. The~prediction that was based on anomalous diffusion that $\alpha_\textmd{L\'evy}^\pi  > \alpha_\textmd{L\'evy}^{\textmd{K}}$ does not seem to be strongly supported. To~clarify this question, further measurements and investigations might be necessary. Considering the intercept parameter $\lambda$, kaons and pions have matching values at around $m_\textmd{T}$ = 0.7 GeV/$c^2$, and~a slight decreasing trend is visible for kaons, possibly due to the worsening of the PID efficiency. The~Lévy-scale R exhibits a similar trend as its Gaussian counterpart; it decreases with $m_\textmd{T}$, and its inverse square is linear in $m_\textmd{T}$, although~this was predicted only for the Gaussian width. A~new empirical scaling variable, $\widehat{R}$, was found, and~it was observed that $\frac{1}{\widehat{R}}$ scales linearly with $m_\textmd{T}$.

\section*{Acknowledgement}
This research was supported by the NKFIH OTKA K-138136 grant.

\end{document}